\documentclass[ amsmath, amssymb,sort&compress ,11pt]{article}
\usepackage{amsmath,amssymb,graphicx}
\usepackage{epstopdf}
\usepackage{hyperref}
\input{epsf.sty}

\newcommand{\ba}{\begin{eqnarray}}
\newcommand{\ea}{\end{eqnarray}}
\newcommand\be{\begin{equation}}
\newcommand\ee{\end{equation}}

\newcommand\cN{{\cal N}}
\newcommand\N{{\mathrm{N}} }

\begin{document}


\title{ Multi-brid DBI Inflation}

\author{Salomeh Khoeini-Moghaddam\\\emph{skheini(AT)khu.ac.ir}\\Department of Astronomy and High Energy Physics,\\ Faculty of Physics, Kharazmi University, Tehran, Iran}

\maketitle

\begin{abstract}
  It is shown that it is possible to apply $\delta \cN$ to some special  non-canonic  cases. We extended the multi-brid idea to the multi-field separable model with a non-canonical kinetic term, mainly the DBI model. Assuming a specific surface for the end of inflation and introducing new fields, enable us to find an explicit expression for the number of e-folds in terms of the new fields. By using $\delta$N formalism, we arrived at the cosmological parameters. We considered the DBI model for two different limits, viz., speed limit and constant sound speed.

{\bf keywords}: early universe; inflation; multi-field; DBI; multi-brid

\end{abstract}


\section{Introduction}
The inflation theory offers an intelligent approach to the basic problems of cosmology\cite{Guth:1980zm,Steinhardt,Linde}. According to this theory, a rapid expansion era in the universe enlarges its physical length  to the nearly 60 times the original size; this rapid expansion solves the old problems encountered in cosmology, such as flatness, horizon and monopole issues.
The fluctuation of the inflaton, the field which drives inflation, induces the fluctuations in the energy density in such a manner that the curvature
power spectrum is nearly scale invariant. There are several cosmological observations which support this theory. Recently, the Planck satellite observed  the Cosmic Microwave Background(CMB) anisotropy and its polarization with a small angular resolution; it results concur with predictions of the inflation theory, in general\cite{Planck 2018a,Planck2018b,Planck2018c}.

 Although  the experimental tests support inflation; from  the theoretical perspective, a proper understanding of the  nature of the factors that produces the inflation is still unclear. In the simplest model, one field,  the inflaton which is minimally coupled to gravity, rolls  very slowly in a very flat potential . The flatness of the potential is essential to achieve enough inflation. One way to relax this condition is to change the dynamics of the inflaton.
When the kinetic term in the Lagrangian of this field is non-canonic the slow-roll condition is not necessary anymore. For example when we have a Dirac-Born-Infeld (DBI) field, there is a fast-roll inflation\cite{Alishahiha:2004eh}. Brane inflation is an example of this model in which the radial distance between a pair of D3- and anti D3-brane takes the role of the inflaton~\cite{dvali-tye,Alexander:2001ks,HenryTye:2006uv,collection,Dvali:2001fw,Kachru:2003sx,Firouzjahi:2003zy,Burgess:2004kv,Buchel,Baumann:2006th,
Baumann:2007ah,Chen:2008au,chen2010,Shandera:2006ax}. In general, it is possible to get inflation from a general class of non-standard kinetic terms, which
is called k-inflation, in this model the inflation can exist even in the absence of the potential \cite{k-inflation,Garriga:1999vw}.
However, the existence of more than one field can relax many limits on  the single field models. In the multi-field models,
co-operation between  the fields can produce inflation, even if each field is not able to drive  the inflation by itself, in such a way that the e-folding number and curvature perturbation are proportional to the number of fields.~\cite{Liddle98,Malik98,n-flation,Staggered,DBI-Nflation,Cai08}.

Regardless of  the model of inflation, this era must end at some time $t_f$. For example, a waterfall mechanism can cease inflation on a specific surface in field space which, is called the end of inflation surface. By using the equation of end of inflation surface, the e-folding number can be expressed explicitly in terms of fields; this concept is known as multi-brid~\cite{sasaki1,sasaki2,book1}. We generalize this idea to the separable non-canonic multi-field models,
assuming a general equation for this surface in terms of fields at $t_f$. Using  $\delta \cN$ formalism, the observational parameters such as spectral index and non-Gaussianity are obtained.

The rest of this paper is organized as follows: in Section \ref{setup} , we present set-up for separable models and use $\delta\cN$ formalism to arrive at observational parameter. In Sections \ref{example} , we apply this method to the multi-speed DBI in speed limit, and DBI with constant sound speed.


\section{The model}\label{setup}
We consider a separable action for N number of scalar fields with  the non-canonical kinetic term, therefore the action is given by
\ba
 S=\frac{1}{2}\int d^4x\sqrt{-g}\hspace{0.5mm}[ M^2_p R+ 2\sum_I^N P_I(X_I,\phi_I) ],
\ea
where $X_I$ is the kinetic term of $\phi_I$, $X_I=-\frac{1}{2}g^{\mu\nu}\nabla_\mu\phi_I\nabla_\nu\phi_I$.

Consider a spatially flat FRW background, $ds^2=-dt^2+a^2(t)d\vec{x}^2$, the variation of action gives the field equations as follows,
\ba\label{eqm1a}
  \frac{d}{dt}(a^3P_{I,X_I}\dot{\phi}_I)-a^3P_{I,\phi_I}=0.
\ea
  As the fluctuation of each field is characterized by its own sound speed, $c_{S_I}=P_{I,X_I}/(P_{I,X_I}+2X_{I}P_{I,X_IX_I})$, this is a "multi-speed" model\cite{Cai09,Pi2012}.

  Rewriting the equations of motion (\ref{eqm1a}) as $\frac{1}{a^3P_{I,\phi}}\frac{d}{dt}\left(a^3P_{I,X}\dot{\phi}_I\right)=1$, leading us to introduce new fields as follows:
\ba
 \ln{q_I}(t)\equiv-\int^{t}{\frac{\dot{a}}{a^4p_{I,\phi_I}}\frac{d}{dt'}\left(a^3P_{I,X_I}\dot{\phi_I}\right)dt'},
\ea
 so according to above definition each $q_I$ is a functional of t,
In terms of these new fields, the equations of motion can be expressed very simply as,
\ba\label{eqm1}
 \frac{d \ln{q_I}}{d t}=-H,
\ea
 where $H=dot{a}/a$ is the Hubble constant. It is obvious that the right hand side of these equations are identical for all the fields; therefore, for each I and J we have,
 \ba
  \frac{ \ln{q_I}}{\ln{q_{J}}}=\frac{\ln{q_{I,f}}}{\ln{q_{J,f}}}.
 \ea
 This indicates that in the space of $q_I$ the motion is radial

For later convenience, the time variable is changed to  the number of e-folds denoting by $\mathcal{N}$ which is defined through
\ba\nonumber
 d\mathcal{N}=-H dt.
 \ea
 The minus sign shows that the number of e-folds is backward in time (at the end of inflation, it vanishes i.e. $\mathcal{N}_f=0$).
From (\ref{eqm1}), in terms of $\cN$, the equation of motion for $q_I$ becomes,
\ba\label{eqm2}
 \frac{d\ln{q_I}}{d\mathcal{N}}=1,
\ea
  which can be solved as
\ba\label{number of e-fold}
 \cN-\cN_f=\ln{q_I}-\ln{q_{I,f}}.
\ea

 It is worth mentioning that there is also another term contributing in N, which is called $\cN_c$. This term stems from the fact that the end of the inflation surface is not a constant energy density surface\cite{sasaki1}. If we assume that immediately after the end of inflation the universe is radiation dominate, it is given by $\cN_c=\frac{1}{4}\frac{\rho_f}{\rho_c}$, where $\rho$ is the energy density; ``f" and ``c" refer to the end of the inflation surface and the surface of constant energy, respectively. Throughout the rest of the paper, we ignore this term.

Suppose the  end of inflation surface is known, for example, a waterfall mechanism terminates the inflation on a specific surface. This surface is determined by a relation between the fields in q space at $\cN_f=0$ as $\mathcal{f}\left(\ln{q_{I,f}}\right)=\mathcal{M}^2$ \cite{sasaki1,sasaki2}.
Throughout the rest of the paper we restrict ourselves to two-field case, allowing us to do some precise calculations. Since the $\ln{q_{I,f}}$s are not independent at the end of the inflation (the end of inflation surface is fixed), in two-field case, there is only one degree of freedom in principle;
denoting by $\theta$
i.e. we can define  $\ln{q_{1,f}}=f_1(\theta)$ and $\ln{q_{2,f}}=f_2(\theta)$.
Variation of (\ref{number of e-fold}) gives,
 \ba\label{variation1}
  \delta\cN=\delta\ln{q_1}-\delta\ln{q_{1,f}}=\delta\ln{q_2}-\delta\ln{q_{2,f}}.
 \ea
Rewriting this equation as
 \ba\label{variation2}
  \delta\ln{q_1}-\delta\ln{q_2}=\delta\ln{q_{1,f}}-\delta\ln{q_{2,f}},
 \ea
  We are interested in perturbation calculation up to second order, so we expand $\ln{q_{I,f}}$  as below:
  \ba\label{variation3}
   \delta\ln{q_{I,f}}=A_I(\delta_1\theta+\delta_2\theta)+A_{II}\frac{(\delta_1\theta)^2}{2},
  \ea
 with $A_I$ and $A_{II}$ being the first and second derivatives of $f_I$ with respect to $\theta$ and also $\delta_1$ and $\delta_2$ refers to first- and second-order variations, respectively.
Replace $\ln{q_{1,f}}$ and $\ln{q_{2,f}}$ in first order perturbation of (\ref{variation2}), $\delta_1\ln{q_1}-\delta_1\ln{q_2}=\delta_1\ln{q_{1,f}}-\delta_1\ln{q_{2,f}}$, we arrive at
 \ba
  \delta_1\theta=\frac{\delta_1\ln q_1-\delta_1\ln q_2}{A_1-A_2}.\nonumber
 \ea
From the second order perturbation of (\ref{variation2}),
$\delta_2\ln{q_1}-\delta_2\ln{q_2}=\delta_2\ln{q_{1,f}}-\delta_2\ln{q_{2,f}}$, we obtain
\ba
 \delta_2\theta&=&\frac{\delta_2\ln q_1-\delta_2\ln q_2}{A_1-A_2}\\\nonumber
 &-&\frac{A_{11}-A_{22}}{2\left(A_1-A_2\right)}\left(\delta_1\theta\right)^2.\nonumber
\ea
Substituting $\delta_1\theta$ and $\delta_2\theta$ in  (\ref{variation3})  gives,
  \ba\label{second orde q}
   \delta_1\ln{q_{1,f}}&=&\frac{A_1}{A_1-A_2}(\delta_1\ln{q_1}-\delta_1\ln{q_2}),\label{first order q}\\
   \delta_2\ln{q_{1,f}}&=&\frac{A_1}{A_1-A_2}\left(\delta_2\ln q_1-\delta_2\ln q_2\right)\\\nonumber
   &+&\frac{1}{2}\frac{A_1A_{22}-A_2A_{11}}{(A_1-A_2)^3}(\delta_1\ln{q_1}-\delta_1\ln{q_2})^2\label{second order q},
  \ea
  there is also a similar expression for $q_{2}$. In terms of the variation of original fields we have                                                 $\delta_1\ln q_I=\frac{\partial\ln q_I}{\partial\phi_I}\delta\phi_I$ and                                                                             $\delta_2\ln q_I=\frac{1}{2}\frac{\partial^2\ln q_I}{\partial\phi_I^2}\left(\delta\phi_I\right)^2$ ( we assume that the variation of original fields is of first order).

\subsection{$\delta\cN$ formalism}
Although the $\delta$N formalism is based on slow-roll assumption\cite{deltaN}, in certain circumstances  it can be applied to separable non-canonical models\cite{beyand,Garriga2016}.
In \cite{Garriga2016} it is shown that utilization of the $\delta$N formalism in the  non-canonical multi-field models, is feasible (for a short review see the appendix.)

Assume that an attractor solution exists, so  that we can apply the $\delta \cN$  formalism to the desired model.
The explicit expression of $\cN$ in terms of $q_I$s (\ref{number of e-fold}) allows us to compute $\delta\cN$ in terms of field's variation. We assume slow variations at the horizon exit to ignore  the $\delta\dot{\phi}_I$. Substituting  $\delta_1\ln q_{I,f}$ in $\delta\cN$  and $\delta\phi_I$ in $\delta\ln{q_I}$s, gives
  \ba\label{delta1 N}
   \delta_1\cN=H\left(\frac{A_2\delta\phi_1/\dot{\phi}_1-A_1\delta\phi_2/\dot{\phi}_2}{A_1-A_2}\right),
  \ea
in the above equation, we have used the relation $\frac{\partial\ln{q_I}}{\partial\phi_I}=\frac{d\ln{q_I}/dt}{d\phi/dt}$ which emerges from the fact that both of $\ln{q_I}$ and $\phi$ are functions of t.
  The second-order variation of e-folding number is as follows:

  \ba
   \delta_2\cN&=&\delta_2\ln q_I-\delta_2\ln q_{I,f}.
  \ea
  Replacing (\ref{second order q}) we arrive at
\ba\label{delta2 N}
\delta_2\cN&=&\frac{1}{2}H^2\frac{A_2A_{11}-A_1A_{22}}{(A_1-A_2)^3}
\left(\frac{(\delta\phi_1)^2}{\dot{\phi}_1^2}+\frac{(\delta\phi_2)^2}{\dot{\phi}_2^2}
-2\frac{\delta\phi_1}{\dot{\phi}_1}\frac{\delta\phi_2}{\dot{\phi}_2}\right)\\\nonumber
&+&\frac{1}{2}\left(\frac{-A_2}{A_1-A_2}\frac{H\ddot{\phi}_1-\dot{H}\dot{\phi}_1}{\dot{\phi}_1^3}\left(\delta\phi_1\right)^2
+\frac{A_1}{A_1-A_2}\frac{H\ddot{\phi}_2-\dot{H}\dot{\phi}_2}{\dot{\phi}_2^3}\left(\delta\phi_2\right)^2\right),
\ea
 in the above equation we  substitute $\left(H\ddot{\phi}_I-\dot{H}\dot{\phi}_I\right)/\dot{\phi}_I^3$ for $\partial^2q_I/\partial \phi_I^2$.
\subsection{Cosmological Parameters}
Assume  that the two-point function of  the scalar field fluctuations at the horizon exit is given by Gaussian distribution,
 \ba
 <\delta\phi_{I}\delta\phi_{J}>_k=\left(\frac{H}{2\pi}\right)^2|_{t_k}\delta_{IJ}
 \ea
where $t_k$ is the horizon crossing time of the co-moving wave-number k, such that  $k\hat{c}_s=Ha$, the $\hat{c}_s$ is the maximum of sound speeds characterized by the final freezing scale.
Using  the $\delta\cN$ formalism the curvature power spectrum and spectral index are given by
\ba
\mathcal{P}_{S}&=&\sum_I\cN_{,I}^2\left(\frac{H}{2\pi}\right)^2\mid_{t_k}
=\left(\left(\frac{H^4}{4\pi^2}\right)\frac{\frac{A_1^2}{\dot{\phi}_1^2}+\frac{A_2^2}{\dot{\phi}_2^2}}{\left(A_1-A_2\right)^2} \right)_{t_k},\\\label{power spectrum}
 n_{s}-1&=&=\frac{d\log{\mathcal{P}_{\zeta}}}{d\log{k}}=-2\epsilon_H+\frac{2}{H}\frac{\sum_{IJ}\dot{\phi}_I\cN_{,I}\cN_{,IJ}}{\sum_K\cN^2_{,K}}\\\nonumber
 &=&-2\epsilon_H-\frac{2}{H^2}\frac{H(A_2^2\ddot{\phi}_1/\dot{\phi}_1^3+A_1^2\ddot{\phi}_2/\dot{\phi}_2^3)-
 \dot{H}(A_2^2/\dot{\phi}_1^2+A_1^2/\dot{\phi}_2^2)}{A_1^2/\dot{\phi}_2^2+A_2^2/\dot{\phi}_1^2},\label{spectral index}
\ea
 where $\cN_{,I}=\partial\cN/\partial\phi_I$ and
  $\epsilon_H$ are defined as usual, as $\epsilon_H=-\frac{\dot{H}}{H^2} $.
Another cosmological parameter is the local non-Gaussianity , given by
\ba
f_{NL}^{local}&=&\frac{5}{6}\frac{\sum_{IJ}\cN_{,I}\cN_{,J}\cN_{,IJ}}{\left(\sum\cN_{,I}^2\right)^2}\\\nonumber
&=&-\frac{5}{6}\frac{A_2A_{11}-A_1A_{22}}{A_1-A_2}\frac{\left(A_1/\dot{\phi}_1^2+A_2/\dot{\phi}_2^2\right)^2}
{\left(\left(A_1/\dot{\phi}_1\right)^2+\left(A_2/\dot{\phi}_2\right)^2\right)^2}\\\nonumber
&+&\frac{5}{6}\frac{A_1-A_2}{H^2}\frac{\left(H\left(A_1^3\ddot{\phi}_2/\dot{\phi}_2^5-A_2^3\ddot{\phi}_1/\dot{\phi}_1^5\right)-
\dot{H}\left(A_1^3/\dot{\phi}_2^4-A_2^3/\dot{\phi}_1^4\right)\right)}{\left(\left(A_1/\dot{\phi}_1\right)^2+\left(A_2/\dot{\phi}_2\right)^2\right)^2}.\label{local non-gaussianity}
\ea
The dependence on $c_s$ is hidden in $\dot{\phi}$ and $\ddot{\phi}$ and the effect of non-trivial end of inflation surface is concealed in $A_I$s. Moreover, there is an equilateral-type non-Gaussianity or a mixed shape because of the non-trivial kinetic term \cite{Cai09,varying-sound-speed,separable,Mixed,Kidani2012}, which we did not calculate explicitly here.

In \cite{sasaki1} each field has its own equation of motion allowing to introduce auxiliary fields. Despite the fact that our model does not satisfy the slow-roll conditions, it is possible to introduce auxiliary fields  because  it is a separable model.These auxiliary fields permit us to continue multi-brid procedure, our results are compatible with \cite{sasaki1}.

\section{Examples}\label{example}
In this section to be more explicit we apply this formalism to two models, DBI in speed limit  and DBI with constant sound speeds. These models are solvable  allowing us to compute cosmological parameters.
\subsection{DBI in speed limit }\label{sec DBI speed limit}
\begin{figure}[t]
\center{
  \includegraphics[width=6cm]{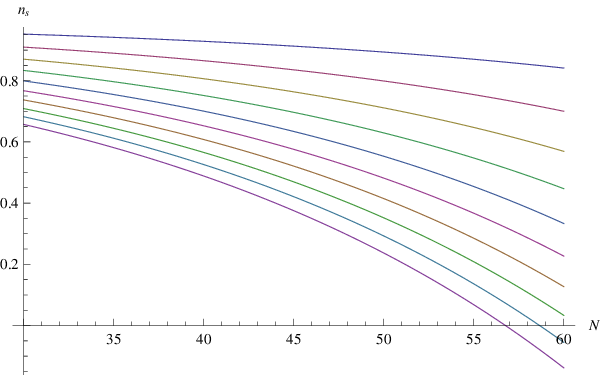}
  \includegraphics[width=6cm]{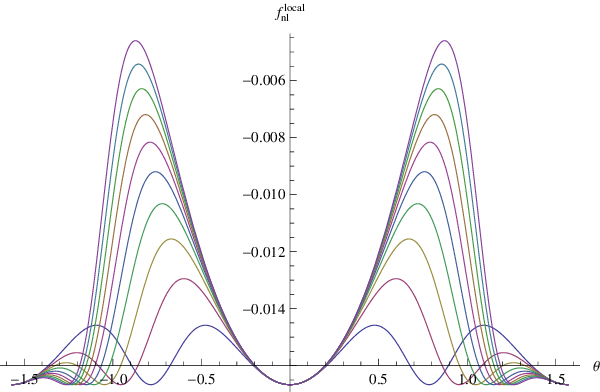}
\caption{ We plot  the spectral index  on the left and the local non-Gassianity on the right for the two-field DBI in the speed limit. We set the ratio of $\lambda_1$ to $\lambda_2$ from 1 to 10. $\theta$ is defined in the tex for parameterizing the end of inflation surface; N is the number of e-folds.}\label{fig1}}
\end{figure}

  Brane inflation, inspired by the string theory, is an example of models with a non-canonical kinetic term. In a simple model, the  distance  between a pair of D3 and anti-D3 brane, which are moving radially in a Calabi-Yau compactification, takes the role of  an inflaton field. This model includes a serious problem; it is  unable to produce enough inflation due to the steepness of the potential between  the brane and anti-brane. To resolve this problem, a warp factor is considered to flatten  the potential, which comes from  the movement of brane and anti-brane inside a warped throat.
  we considered two stacks of coinciding brane, inside a warped throat at different places, moving  ultra relativistically towards the anti-branes  located at the bottom of the throat. The distances between  the two brane stacks of and  the anti-branes take on the role of the inflaton; therefore, this is a
  two-field model \cite{firouz1011,Battefeld:2010rf}. We assume that the background metric is a flat FRW; the action,  therefor is as follows:
\ba\label{dbi action}
 S=\frac{1}{2}\int d^4x\sqrt{-g}\left(M^2_{pl}R+2\sum_I[f_I^{-1}\left(1-\sqrt{1-f_I\dot{\phi}_I^2}\right)-V(\phi_I)]\right),
 \ea
 with  $f_I=\frac{\lambda_I}{\phi_I^4}$. $\lambda_I$ is proportional to the number of brane in each stack. We define {\it{Lorentz factor}} for each field as$\gamma_I=\frac{1}{\sqrt{1-f_I\dot{\phi_I}^2}}$  which is the inverse of the sound speed for each field. The equation of motion is given by
\ba\label{eqmdbi}
\ddot{\phi}_I+3H\gamma_I^{-2}\dot{\phi}_I+\frac{3}{2}\frac{f'_I}{f_I}\dot{\phi}^2_I-\frac{f'_I}{f_I^2}+\gamma_I^{-3}\left(V'_I+\frac{f'_I}{f_I^2}\right)=0
\ea
 In the speed limit  $\gamma_I\gg1$ and $\dot{\phi_I^2}\approx f_I^{-1}$, this equation reduces to
 $\ddot{\phi}_I+\frac{3}{2}\frac{f'_I}{f_I}\dot{\phi}_I^2-\frac{f'_I}{f_I^2}=0,$
which can be solved as $\phi_I(t)\approx\frac{\sqrt{\lambda_I}}{t}$, replacing in the Friedmann equations to get the Hubble constant as
$ H=\frac{\Lambda}{t}$,
where $\Lambda^2=\sum_I(\frac{\lambda_I m_I^2}{6M_P^2})$.
 Plugging this solution back into (\ref{eqmdbi}), up to the next leading correction, the solution is as follows:
\ba\label{solutiondbi}
 \phi_I(t)\simeq\frac{\sqrt{\lambda_I}}{t}\left(1-\frac{9H^2}{2m^4_It^2}\right),
\ea
once again replaced in the Friedmann equation gives the Hubble constant,
$H=\frac{\Lambda}{t}\left(1-\frac{\left(\Delta/\Lambda\right)^2}{t^2}\right)$, with$\Delta^2=\frac{\sum_I\alpha_Im_I^2\lambda_I}{6M_P^2}$. For the sake of simplicity, we introduce the small parameters, $\alpha_I=\frac{9H^2}{2m_I^4}$  and $\alpha=\frac{\Delta^2}{\Lambda^2}$. The e-folding number is obtained through the integration of H with respect to time,
 \ba\label{e-fold dbi corrected}
  \cN&=&\Lambda\left(-\ln{t}-\frac{\alpha}{2t^2}\right)-\Lambda\left(-\ln{t_f}-\frac{\alpha}{2t_f^2}\right).
 \ea
The equation given above enables us to designate the new fields as
\ba
 \ln{q_I}= -\Lambda\ln{t}-\frac{\Lambda}{2}\frac{\alpha}{t^2}.
\ea
Assume that the surface of  the end of inflation is, $g_1^2\phi_{1,f}^2+g_2^2\phi_{2,f}^2=\sigma^2$,
with $\sigma$ being a constant;  up to  the first order in  the small parameters, we have,
\ba\nonumber
 A_1&=&-\Lambda\tan\theta\left(1+\frac{\alpha\sigma^2}{g_1^2\lambda_1}\cos^2\theta\right),\\\nonumber A_2&=&+\Lambda\cot\theta\left(1+\frac{\alpha\sigma^2}{g_2^2\lambda_2}\sin^2\theta\right),\\\nonumber
 A_{11}&=&-\Lambda\left(1+\tan^2\theta\right)\left(1-\frac{\alpha\sigma^2}{g_1^2\lambda_1}\sin 2\theta\right)\\\nonumber A_{22}&=&-\Lambda\left(1+\cot^2\theta\right)\left(1+\frac{\alpha\sigma^2}{g_2^2\lambda_2}\sin 2\theta\right).
 \ea
 With $\tan\theta=g_2\sqrt{\lambda_2}/\left(g_1\sqrt{\lambda_1}\right)$. The power spectrum, spectral index, and local non-Gaussianities are as follows:
\ba
P_{S}&=&\frac{\Lambda^4}{4\pi^2\lambda_1\lambda_2}\left(\lambda_1\sin^4\theta+\lambda_2\cos^4\theta\right)|_{t_k}+\mathcal{O}(\alpha,\alpha_I),\\
n_{S}&=& 1-\frac{4\alpha }{\Lambda t^2}m\\
f_{NL}^{local}&=&\frac{5}{3\Lambda}\frac{\left(-\lambda_1\tan\theta+\lambda_2\cot\theta\right)^2}{\left(\lambda_1\tan^2\theta+\lambda_2\cot^2\theta\right)^2}\\
&-&\frac{5}{6\Lambda}\left(\tan\theta+\cot\theta\right)
\frac{\lambda_1^2\tan^3\theta+\lambda_2^2\cot^3\theta}{\left(\lambda_1\tan^2\theta+\lambda_2\cot^2\theta\right)^2}+\mathcal{O}(\alpha,\alpha_I),
\ea

 as before, $t_k$ is the time of horizon crossing corresponding to the maximum of the sound speed. As the leading terms in spectral index cancel each other, we need to compute the first-order term. Up to zeroth-order of the small parameters, the $P_S$  and $f_{nl}^{local}$ have no H dependance. Assuming $m_1\sim m_2\sim10^{-5}M_p$ and using $\Lambda\sim50$ (which comes from  the COBE normalization for  the curvature power spectrum, $P_S\sim2\times10^{-9}$) we get $\lambda_1\sim\lambda_2\sim10^{14}$. By appropriately choosing  model parameters such as $\Lambda$ and $\lambda_I$ ; observational parameters lie within the Planck range \cite{Planck2018b,Planck2018c} (see Fig (\ref{fig1})).

\subsection{ Two field DBI with constant sound speeds}\label{constant sound speed}
\begin{figure}[h]
 \center{
  \includegraphics[width=6cm]{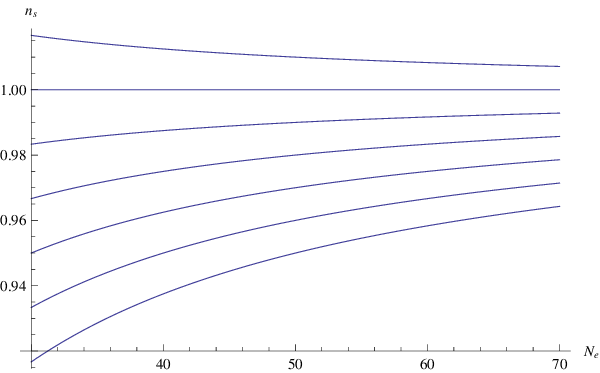}
 \caption{The spectral index  for two-field DBI  with constant sound speed is plotted versus the number of e-folds  $N_e$ for different values of q, the parameter q goes from -3 to 3. }\label{fig2-1}}
\end{figure}

\begin{figure}[h]
 \center{
 \includegraphics[width=3.7cm]{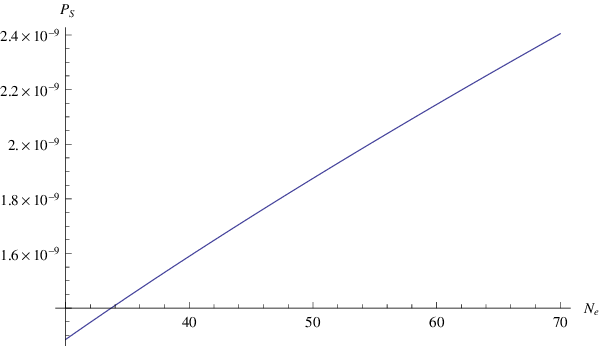}
 \includegraphics[width=3.7cm]{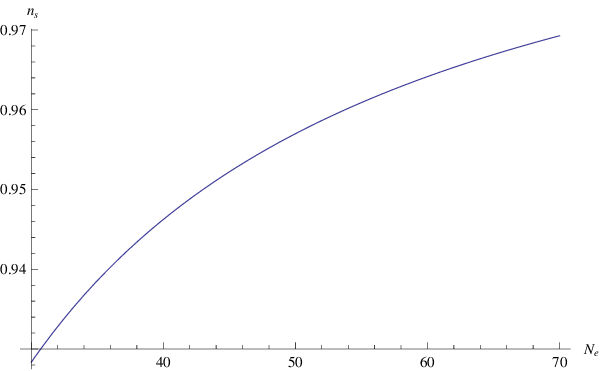}
 \includegraphics[width=3.7cm]{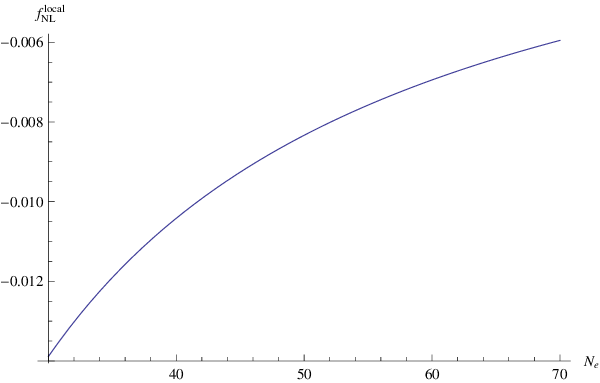}
 \caption{ The power spectrum, spectral index and local non-Gaussianity  for two-field DBI  with constant sound speed  is depicted  versus the number of e-folds. q is selected as -2.3. The other parameters are  $v_1 = 5\times{10^-12},v_2 = 10^-12, g_1=g_2=3\times10^{-9}, c_1=0.1$ and  $c_2=0.2$.}\label{fig2-2}}
\end{figure}

In \cite{copland 2010}, it is observed that for a single-field DBI model
 with $f(\phi)=f_0\phi^{q+2}$, when the potential has a special form as $V(\phi)=V_0\phi^{-q}$, there is an attractor solution with a constant sound speed, $c_s=\sqrt{3/\left(16f_0V_0+3\right)}$. An extension of this solution  with two fields is investigated in \cite{Kidani2012}.
  We extend this kind of solution to multi-speed case. Similar to the previous Section, we consider two stacks containing $p_1$ and $p_2$ branes moving towards $p_1+p_2$  anti-branes at the bottom of the warped throat. We are interested in the potentials that include attractor solutions with constant sound speed, $V_I=V_{0,I}\phi_I^{-q}$. The form of the action is similar to (\ref{dbi action}) with $f_I\left(\phi\right)=f_0\phi_I^{q+2}$, where the equation of motion is rewritten as,
  \ba\label{eqm cspeed}
   \ddot{\phi}_I+3H\dot{\phi}_I-\frac{\dot{c}_{sI}}{c_{sI}}+c_{sI}V'_{I}-\frac{\left(1-c_{sI}\right)^2}{2}\frac{f'_I}{f^2_I}=0.
  \ea
  This equation has the solution as $c_{sI}=\sqrt{3/\left(q^2f_0V_0+3\right)}$ and $\phi_I=\phi_{I,0}t^{\frac{2}{4+q}}$ with $\phi_{0,I}=\left(\sqrt{\frac{1-c^2_{s,I}}{f_0}}\frac{2}{4+q}\right)^{\frac{2}{4+q}}$. Substituting in the Friedmann equations and keeping only the leading terms in late time, we arrive at
  \ba\label{Hubble cspeed}
    H&=&\tilde{\Lambda}t^{-\frac{q}{4+q}},
  \ea
   where we define $\tilde{\Lambda}^2=\frac{1}{3M^2_p}\left(\frac{4+q}{2}\right)^{\frac{2}{4+q}}\left(\frac{\left(\sum_IV_{0,I}(1-c^2_{s,I})\right)}{f_0}\right)^{\frac{1}{4+q}}$.
  Integration of H give us the number of the e-folding,
  \ba\label{e-fold cspeed}
   \cN=-\tilde{\Lambda}\frac{4+q}{4}\left(t^{\frac{4}{4+q}}-t_f^{\frac{4}{4+q}}\right),
  \ea
 in which, as before, we select $\cN_f=0$. On comparing with (\ref{number of e-fold}), it directs us to define new fields as,
  \ba
   \ln{Q_I}=-\frac{\tilde{\Lambda}\left(4+q\right)}{4}t^{\frac{4}{4+q}}.
  \ea
  Assume that the inflation ends when $g_1^2\phi_{1,f}^2+g_2^2\phi_{2,f}^2=\sigma^2$, we have
  \ba
  A_1&=&-\frac{\beta_1\sigma^2}{g_1^2}\sin{2\theta},\hspace{1cm}A_2=\frac{\beta_2\sigma^2}{g_2^2}\sin{2\theta}\\\nonumber
  A_{22}&=&-2\frac{\beta_1\sigma^2}{g_1^2}\cos{2\theta},\hspace{0.6cm}A_{22}=2\frac{\beta_1\sigma^2}{g_1^2}\cos{2\theta},
  \ea
with $\beta_I=-\frac{4+q}{4}\tilde{\Lambda}\frac{1}{\phi_{0,I}^2}$ and $\tan\theta=\frac{g_2\phi_{2,f}}{g_1\phi_{1,f}}$. Replacing in (\ref{power spectrum}),(\ref{spectral index}) and (\ref{local non-gaussianity}) gives,
\ba
P_S&=&-\frac{\left(4+q\right)\tilde{\Lambda}^3}{4\pi^2}\left(-\frac{4+q}{4}\right)^{\frac{q}{2}+1}\left(\frac{\tilde{\Lambda}}{\cN}\right)^{\frac{q}{2}+1}
\frac{\beta_1\beta_2\left(\beta_1g_2^4+\beta_2g_1^2\right)}{\left(\beta_1g_2^2+\beta_2g_1^2\right)^2},\\\label{power cspeed}
n_S&=&1-\frac{2-q}{2\cN}\\\label{spectral cspeed}
f_{nl}^{local}&=&-\frac{5}{12}\frac{\left(\beta_1g_2^2+\beta_2g_1^2\right)\left(\beta_1g_2^6+\beta_2g_1^6\right)}
{\beta_2g_1^4+\beta_1g_2^4}\frac{1}{\cN}\label{non-gaussianity cspeed}
\ea
 Neither of the cosmological parameters depends on $\theta$, which is characterized the surface of the end of inflation. The Figure (\ref{fig2-1}) depicts $n_s$ versus the number of e-folds,to be consistent with  Plank data we select $q=-2.3$ which corresponds to $f=f_0\phi^{-0.3}$ and $V=V_0\phi^{2.3}$, then  we plot the other cosmological parameters(Fig.\ref{fig2-2}).

\section{Conclusion}
\label{conclusion}
In this paper, we extended "multi-brid" inflation concept to a multi-field separable models. Multi-brid idea is based on $\delta \cN$ formalism  which depends on slow-roll conditions. However it is shown that it is possible to apply $\delta \cN$ formalism to separable models\cite{Garriga2016}, to be more precise we review the the formalism in the appendix. In non-canonic separable models each field is minimally coupled to other fields in such a way as each field has its own equation of motion, this independence from other fields permits us to define auxiliary fields which provide a straightforward way to deal with the model. We assume that the end of inflation surface is known. By equating the variations, order by order, we found the variation at the end of inflation surface in terms of the variations in the fields and replaced it in the e-fold variations. Using $\delta\cN$ formalism, we gain the observational parameters such as scalar spectral index, its power spectrum and local non-Guassianity. Our results are compatible with previous work \cite{sasaki1}.
By applying this set-up to the multi-speed DBI model in the speed limit and DBI with constant sound speed, we obtain the cosmological parameters for these models.

\vspace{1cm}
\noindent\textbf{Acknowledgments}

{The author would like to thank H. Firouzjahi for instructive discussions and useful comments.}



\setcounter{equation}{0}
\appendix
\section{Development of $\delta\cN$ formalism }
In this section we review the approach of \cite{Garriga2016} to extend $\delta\cN$ formalism  to more general cases by using a super-potential formalism.
\subsection{The background}
 The Friedmann equations are,
\ba
H^2&=&\frac{\kappa^2}{3}\left(2\Sigma_IP_{I,X_I}X_I-P_I\right)\\\label{fried1}
\dot{H}&=&-\frac{\kappa^2}{2}\Sigma_IP_{I,X_I}\dot{\phi}_I\label{fried2}
\ea
where $\kappa^2=8\pi G$. We can define momentum as
\ba
\pi\equiv P_{I,X_I}\dot{\phi}_I
\ea
Since the Lagrangian is not singular, it is possible to obtain $\dot{\phi}_I$ as a function of $\phi_I$ and $\pi_I$ i.e.
\ba
\dot{\phi}_I=F_I\left(\phi_I,\pi_I\right).
\ea
In principle, it is possible to solve the equations of motion and gain  $\pi_I$ in terms of $\phi_I$ and initial momenta,$c_I$s.

Now we can replace $\dot{\phi}_I$ in energy density and express Hubble Parameter as below,
\ba\label{hubble1}
H=2W\left(\phi_J,c_J\right)
\ea
From(\ref{fried2}) and (\ref{hubble1}) it is reasonable to conclude that
\ba
\pi_I=-\frac{4}{\kappa^2}\frac{\partial W}{\partial\phi_I}
\ea

Replacing in (\ref{fried1}), gives a differential equation for W as,
\ba
W^2=\frac{\kappa^2}{12}\rho[\phi_I,\frac{\partial W}{\partial\phi_I}]
\ea
"W" is called superpotential.
\subsection{$\delta\cN$ in non slow-roll limit}
The $\delta\cN$ formalism presume the separate universe assumption. The separate universe approach states that when the the physical scale of fluctuations, L, in comparison with the Hubble length, $H^{-1}$, is very big i.e.$L\gg H^{-1}$ each region of Hubble size can be considered as a FRW universe; therefore each patch evolves independently. We define small parameter $\varepsilon$ as $\varepsilon\equiv\frac{1}{LH}$ and expand the equations in the order of that parameter.
We employ ADM formalism,
\ba\label{ADM}
ds^2=g_{\mu\nu}dx^\mu dx^\nu-\alpha dt^2+\gamma_{ij}\left(dx^i+\beta^idt\right)\left(dx^j+\beta^jdt\right)
\ea
where $\alpha$ and $\beta^i$ are lapse and shift vector respectively. $\gamma^{ij}$ is spatial metric which for later convenience decomposed as,
\ba
\gamma_{ij}=a^2e^{2\mathcal{R}\left(x\right)}[e^{h\left(x\right)}]_{ij}
\ea
$tr[h]=0$. As usual the extrinsic curvature is defined as below,
\ba\label{extrinsic}
K=\nabla_\mu n^\mu=\frac{1}{\sqrt{-g}}\partial_\mu\left(\sqrt{-g}n^\mu\right)
\ea
the $n^\mu$ is unit tangent vector to time-like congruence orthogonal to t=costant hyper-surfaces, $n^\mu=\alpha^{-1}\left(1,-\beta^i\right)$, replacing in (\ref{extrinsic}) we obtain,
\ba\label{aextrinsic}
K=\alpha^{-1}[3\left(H+\dot{\mathcal{R}}\right)-D_i\beta^i]
\ea
 The the expansion along these normal congruence can be interpreted as difference in number of e-folding, therefore we define the number of e-folds as below,
 \ba\label{adela N}
 \N=\frac{1}{3}\int\alpha K dt.
 \ea
The derivative with respect to $\N$ is defined as,
\ba
\partial_\N\equiv\frac{3}{\alpha K}\partial_t
\ea
It is assumed that in the limit $\varepsilon\rightarrow0$ we arrive at FRW. We choose the gauge in which
\ba
\partial^ih_{ij}=0.
\ea

In gradient expansion we have\cite{deltaN},
\ba
\partial_j\beta^i&=&\mathcal{O}\left(\varepsilon^2\right)\\
\dot{h}_{ij}&=&\mathcal{O}\left(\varepsilon^2\right)
\ea

and for scalar fields
\ba
T_{ij}-\frac{1}{3}\gamma^{kl}T_{kl}\gamma_{ij}=\mathcal{O}\left(\varepsilon^2\right)
\ea

Up to second order 0f $\varepsilon$, the Einstein equations and the field equations of the scalar fields read as\cite{beyand},
\ba
K^2&=&\frac{\kappa^2}{3}\left(K^2P_{I,X_I}\partial_\cN\phi_I\partial_\cN\phi_I-9P\right)+\mathcal{O}\left(\varepsilon^2\right),\label{per1}\\
\partial_\N K&=&-\frac{\kappa^2}{2}KP_{I,X_I}\partial_\N\phi_I\partial_\N\phi_I+\mathcal{O}\left(\varepsilon^2\right),\label{per2}
\ea
\ba
K\partial_\N\left(KP_{I,X_I}\partial_\N\phi_I\right)&+&3K^2P_{I,X_I}\partial\N\phi_I-9P_{I,\phi_I} =\mathcal{O}\left(\varepsilon^2\right),\label{per3}
\ea
and momentum constraint is as follows,
\ba
\partial_iK=-\frac{\kappa^2}{2}KP_{I,X_I}\partial_\N\phi^I\partial_i\phi^I+\mathcal{O}\left(a\varepsilon^3\right).\label{per4}
\ea
By taking the spatial derivative of (\ref{per1}) and replacing (\ref{per2}) and (\ref{per3}), we arrive at
\ba
\partial_iK=-\frac{\kappa^2}{2}KP_{I,X_I}\partial_\N\phi_I\partial_i\phi_I+B_i+\mathcal{O}\left(a\varepsilon^3\right),\label{per5}
\ea
where
\ba\label{B_i}
B_i=\frac{\kappa^2K}{2\partial_\N\ln{e^{3\N}}K}[\partial_\N\phi_I\partial_i\left(P_{I,X_I}\partial_\N\phi_I\right)-\partial_\N\left(P_{I,X_I}\partial_\N\phi_I\right)\partial_i\phi_I].
\ea
By comparing (\ref{per4}) and (\ref{per5}), it is obvious that for ensuring consistency between Hamilton and momentum constraints, we must have
\ba
a^{-1}B_i=\left(\varepsilon^3\right)
\ea
In \cite{Garriga2016} it is shown that under attractor assumption, this condition is satisfied in more general case than slow-roll condition. With the same argument we said in the background, it is possible to expand the field equation in terms of superpotential as,
\ba
P_{I,X_I}\partial_\N\phi^I=-\frac{2}{\kappa^2}\frac{\partial\ln W}{\partial\phi^I}+\mathcal{O}\left(a\varepsilon^2\right)
\ea
where $K=6W+\mathcal{O}\left(a\varepsilon^2\right)$. In attractor regime,  we can ignore the dependence of $W$ on $c_I$s and the leading term in (\ref{B_i}) vanishes so the consistency condition is satisfied.

From the above discussion we have
\ba
D_i\beta^i= \mathcal{O}\left(\varepsilon^2\right),
\ea
so we can ignore the last term in (\ref{aextrinsic}).
We define
\ba
\delta\cN\left(t_2,t_1;x\right)\equiv\N\left(t_2,t_1;x\right)-\cN\left(t_2,t_1\right)
\ea
in which $\cN\left(t_2,t_1\right)$ is e-folding number in unperturbed universe, replacing (\ref{aextrinsic}) in (\ref{adela N}) and neglecting the terms of order $\mathcal{O}\left(\varepsilon^2\right)$ we arrive at
\ba
\delta\cN\left(t_2,t_1;x\right)=\mathcal{R}\left(t_2,x\right)-\mathcal{R}\left(t_1,x\right)
\ea
At initial space-like hyper-surface $\Sigma_i$ the spatial curvature vanishes, i.e. $\mathcal{R}\left(t_i,x\right)=0$. We choose a constant energy density hyper-surface,$\Sigma_f$ ,as the final hyper-surface $\rho\left(t_f,x\right)\equiv\rho_f=constant$. By a appropriate choice of coordinate system, these two hyper-surfaces can be made to be constant t slices. We define the adiabatic curvature perturbation as,
\ba
\zeta\left(t_f,x\right)\equiv\mathcal{R}\left(t_f,x\right)
\ea
Setting $t_1=t_i$ and $t_2=t_f$ and using separate universe approach we arrive at the usual relation between curvature perturbation and $\delta\cN$,
\ba
\zeta\left(t_f,x\right)=\delta\cN\left(t_f,t_i;x\right)\equiv\delta\cN\left(t_f;\delta\phi_i\left(x\right)\right)
\ea
in the above equation we assume that $\delta\phi_i\left(x\right)\equiv\delta\phi\left(t_i,x\right)$ doesn't depend on momentums.

\begin{thebibliography}{99}
\vspace*{-1mm}
\begin{small}\baselineskip=10pt\itemsep-2pt
\bibitem{Guth:1980zm}
   A.~H.~Guth,
   ``The Inflationary Universe: A Possible Solution To The Horizon And Flatness Problems,''
    Phys.\ Rev.\ D {\bf 23}, 347 (1981);

\bibitem{Linde}
  A.~D.~Linde,
   ``A New Inflationary Universe Scenario: A Possible Solution Of The Horizon,
   Flatness, Homogeneity, Isotropy And Primordial Monopole Problems,''
   Phys.\ Lett.\ B {\bf 108}, 389 (1982);

\bibitem{Steinhardt}
   A.~Albrecht and P.~J.~Steinhardt,
   ``Cosmology For Grand Unified Theories With Radiatively Induced Symmetry Breaking,''
   Phys.\ Rev.\ Lett.\  {\bf 48}, 1220 (1982).

\bibitem{Planck 2018a}
        Planck Collaboration ,
      ``Planck 2018 results. VI. Cosmological parameters,''
      [arXiv::1807.06209]

\bibitem{Planck2018b}
      Planck Collaboration ,
      ``Planck 2018 results. X. Constraints on inflation''
      [arXiv:1807.06211].

\bibitem{Planck2018c}
    Planck Collaboration,
   ``Planck 2018 results. IX. Constraints on primordial non-Gaussianity''
   [arXiv:1905.05697].

\bibitem{Alishahiha:2004eh}
  M.~Alishahiha, E.~Silverstein and D.~Tong,
  ``DBI in the sky,''
  Phys.\ Rev.\  D {\bf 70}, 123505 (2004),
  [arXiv:hep-th/0404084].

\bibitem{dvali-tye}
G.~Dvali and S.-H.H.~Tye,
"Brane Inflation",
Phys. Lett. {\bf B450} (1999) 72,
[hep-ph/9812483].

\bibitem{Alexander:2001ks}
  S.~H.~S.~Alexander,
  ``Inflation from D - anti-D brane annihilation,''
  Phys.\ Rev.\  D {\bf 65}, 023507 (2002)
  [arXiv:hep-th/0105032].

\bibitem{HenryTye:2006uv}
  S.~H.~Henry Tye,
  ``Brane inflation: String theory viewed from the cosmos,''
  Lect.\ Notes Phys.\  {\bf 737}, 949 (2008)
  [arXiv:hep-th/0610221].

\bibitem{collection}
  C.~P.~Burgess, M.~Majumdar, D.~Nolte, F.~Quevedo, G.~Rajesh and R.~J.~Zhang,
  ``The Infationary Brane-Antibrane Universe''
   JHEP {\bf 07} (2001) 047,
 [hep-th/0105204].

\bibitem{Dvali:2001fw}
  G.~R.~Dvali, Q.~Shafi and S.~Solganik,
 ``D-brane inflation,''
  [hep-th/0105203].

\bibitem{Kachru:2003sx}
S.~Kachru, R.~Kallosh, A.~Linde, J.~Maldacena, L.~McAllister
and S.~P.~Trivedi,
`` Towards inflation in string theory'',
JCAP {\bf 0310} (2003) 013,
 [hep-th/0308055].

\bibitem{Firouzjahi:2003zy}
H.~Firouzjahi and S.-H.~H.~Tye,
``Closer towards inflation in string theory,''
Phys.\ Lett.\ B {\bf 584}, 147 (2004),
[hep-th/0312020].

\bibitem{Burgess:2004kv}
C.~P.~Burgess, J.~M.~Cline, H.~Stoica and F.~Quevedo,
 ``Inflation in realistic D-brane models,''
 JHEP {\bf 0409}, 033 (2004),
 [hep-th/0403119].

\bibitem{Buchel}
  A.~Buchel and R.~Roiban,
  ``Inflation in warped geometries,''
  Phys.\ Lett.\ B {\bf 590}, 284 (2004)
  [arXiv:hep-th/0311154].

\bibitem{Baumann:2006th}
  D.~Baumann, A.~Dymarsky, I.~R.~Klebanov, J.~M.~Maldacena, L.~P.~McAllister and A.~Murugan,
  ``On D3-brane potentials in compactifications with fluxes and wrapped D-branes,''
  JHEP {\bf 0611}, 031 (2006)
  [arXiv:hep-th/0607050].

\bibitem{Baumann:2007ah}
  D.~Baumann, A.~Dymarsky, I.~R.~Klebanov and L.~McAllister,
  ``Towards an Explicit Model of D-brane Inflation,''
  JCAP {\bf 0801}, 024 (2008)
  [arXiv:0706.0360].

\bibitem{Chen:2008au}
  F.~Chen and H.~Firouzjahi,
  ``Dynamics of D3-D7 Brane Inflation in Throats,''
  JHEP {\bf 0811}, 017 (2008)
  [arXiv:0807.2817].

\bibitem{chen2010}
  H.Y.~Chen, J.k.~Gongb, K.~Koyamac and G.~Tasinatod,
 ``Towards multi-field D-brane inflation in a warped throat''
  JCAP {\bf11}(2010)034
  [arXiv:1007.2068]

\bibitem{Shandera:2006ax}
  S.~E.~Shandera and S.~H.~Tye,
  ``Observing brane inflation,''
  JCAP {\bf 0605}, 007 (2006)
  [arXiv:hep-th/0601099].

\bibitem{k-inflation}
 C.~Armendariz-Picon, T.~Damour, V.~Mukhanov,
 `` K-inflation,''
 Phys.\ Lett.\ {\bf B458},209-218(1999),
 [arXiv:hep-th/9904075].

\bibitem{Garriga:1999vw}
  J.~Garriga and V.~F.~Mukhanov,
  ``Perturbations in k-inflation,''
  Phys.\ Lett.\  B {\bf 458}, 219 (1999),
  [arXiv:hep-th/9904176].

\bibitem{Liddle98}
   A.R.~Liddle, A.~Mazumdar and F.E.~Schunck,
   ``Assisted inflation,''
   Phys.\ Rev.\ D{\bf58}, 061301(1998),
   [arXiv:astro-ph/9804177].

 \bibitem{Malik98}
  K.~A.~Malik and D.~Wands,
  ``Dynamics of Assisted Inflation'',
   Phys.Rev.{\bf D59}(1999), 123501,
  [arXiv:astro-ph/9812204].

\bibitem{n-flation}
   S.~Dimopoulos, S.~Kachru, J.~McGreevy, J.~Wacker,
  ``N-flation,''
  JCAP {\bf 0808},003(2008),
  [arXiv:hep-th/0507205].

\bibitem{Staggered}
   D.~Battefeld, T.~Battefeld and A.~C.~Davis,
  ``Staggered Multi-Field Inflation,''
  JCAP {\bf 0810}, 032 (2008)
  [arXiv:0806.1953 ].

\bibitem{DBI-Nflation}
  J.~Ward,
  ``DBI N-flation,''
  	JHEP{\bf 0712},045(2007),
  [arXiv:0711.0760].

 \bibitem{Cai08}
    Y.F.~Cai and W.~Xue,
    ``N-flation from multiple DBI type actions,''
    Phys .\ Lett.{\bf B680},395-398(2009),
    [arXiv:0809.4134].

\bibitem{sasaki1}
    M.~Sasaki,
    ``Multi-brid inflation and non-Gaussianity,''
     Prog.\ Theor.\ Phys.\ {\bf120},159-174(2008),
    [arXiv:0805.0974].

\bibitem{sasaki2}
   A.~Naruko and M.~Sasaki,
   ``Large non-Gassianity from multi-brid inflation,''
   Prog.\ Theor.\ Phys.\ {\bf 121},193-210(2009),
   [arXiv:0807.0180].

\bibitem{deltaN}
N.S. Sugiyama, E. Komatsu and T. Futamase,`` $delta$N formalism'',
 Phys. Rev. D {\bf87} (2013) 023530
[arXiv:1208.1073]

\bibitem{book1}
A.~Abolhasani, H.~Firouzjahi, A.~Naruko, M.~Sasaki , ``Delta N Formalism in Cosmological Perturbation Theory''
World Scientific Publishing Co,2019.

\bibitem{beyand}
 A.~Naruko, Y.~ Takamizu and M.~Sasaki,
 ``Beyond $\delta$N formalism''
 Prog. Theor. Exp. Phys\ {043E01 },(2013)
 [arXiv:]

\bibitem{Garriga2016}
 J.~Garriga, Y.~Urakawa and F.~Vernizzie,
 ``$\delta$N formalism from superpotential and holography''
  JCAP {\bf 02}, 036 (2016)
  [arXiv:1509.07339]

\bibitem{Cai09}
    Y.-F.~Cai and H.-Y.~Xia,
    ``Inflation with multiple sound speeds: a model of multiple DBI type actions and non-Gaussianities'',
    Phys. Lett.{\bf B 677}, 226-234(2009),
    [arXiv:0904.0062],

\bibitem{Pi2012}
    S.~Pi and D.~Wang,
    ``Dynamics of Cosmological Perturbations in Multi-Speed Inflation'',
    Nucl.\ Phys.\ {\bf B862}(2012)409,
    [arXiv:1107.0813].

\bibitem{varying-sound-speed}
   J.~Emery, G.~Tasinato and D.~Wands,
 ``Local non-Gaussianity from rapidly varying sound speeds''
  JCAP{\bf 08}(2012)005,
  [arXiv:1203.6625].

\bibitem{separable}
   A.~Mazumdar, L.~Wang,
 ``Separable and non-separable multi-field inflation and large non-Gaussianity,''
 JCAP{\bf09}(2012)005,
   [arXiv:1203.3558].

\bibitem{Mixed}
   J.~Emery, G.~Tasinato, D.~Wands,
  ``Mixed non-Gaussianity in multiple-DBI inflation,''
	JCAP{\bf05}(2013)021,
   [arXiv:1303.3975].

\bibitem{Kidani2012}
 T.~Kidani, K.~Koyamabk, and S.~ Mizuno
 ``Non-Gaussianities in multi-field DBI inflation with a waterfall phase transition''
 Phys.\ Rev.\ D{\bf86} (2012) 083503
 [arXiv:1207.4410]

\bibitem{firouz1011}
   H. Firouzjahi and S. khoeini-Moghaddam;
   ``Fields Annihilation And Particles creation in DBI inflation,''
   JCAP{\bf1102}(2011)012,
   [arXiv:1011.4500].

\bibitem{Battefeld:2010rf}
  D.~Battefeld, T.~Battefeld, H.~Firouzjahi and N.~Khosravi,
  ``Brane Annihilations during Inflation,''
  JCAP {\bf 1007}, 009 (2010)
  [arXiv:1004.1417 [hep-th]].

\bibitem{copland 2010}
E.J.~Copeland, S.~Mizuno and M.~Shaeri
``Cosmological Dynamics of a Dirac-Born-Infeld field,''
Phys.\ Rev.\ D{\bf 81}(2010) 123501,
[arXiv:1003.2881]

\end{small}
\end{thebibliography}
\end{document}